\documentstyle[twocolumn,prl,aps]{revtex}
\begin{document}
\draft 
%\wideabs{
\title{ Relativity of Entanglement } 
\author{ Won-Young Hwang\cite{email}, Jinhyoung Lee\cite{byline},
  Doyeol (David) Ahn\cite{byline2}, 
 and Sung Woo Hwang\cite{byline3}}
\address{ Institute of Quantum Information Processing and Systems,
  University of Seoul, 90 Jeonnong, Tongdaemoon, Seoul 130-743,
  Korea }
\maketitle
\begin{abstract}
It has recently been suggested that various entanglement measures
for bipartite mixed states do not in general give the same ordering 
even in the asymptotic cases
[S. Virmani and M. B. Plenio, Phys. Lett. A {\bf 268}, 31 (2000)].
That is, for two certain mixed states, the order of the
degree of entanglement depends on the measures. 
Therefore, incomparable pairs of
mixed states which cannot be transformed to each other with unit
efficiency by any combinations of local quantum operations and
classical communications exist. 
We make an analogy of the relativity of the order of the degree of
entanglement 
to the relativity of temporal orders in the special theory of relativity.
\end{abstract}
%}

%\section{Introduction}
%\label{sec:level1}

Quantum entanglement led to a controversy over the
Einstein-Podolsky-Rosen experiment \cite{eins} and to the 
nonlocality of
quantum mechanics \cite{bell}.
On the other hand, entanglement is one of the
key ingredients in quantum information processing
\cite{kim}. For example, 
a speedup in quantum computation \cite{shor} is obtained
through 
parallel quantum operations on massively superposed states,
which are, in general, entangled.

Recently, it has been shown that different entanglement measures can
give rise to different orderings for mixed states of bipartite systems
\cite{virm,eise}. In this paper, we make an analogy between the
non-uniqueness (or relativity) of the order of the degree of 
entanglement and the
relativity of temporal order in the special theory of relativity
\cite{tayl}. The temporal order of two certain events can be
reversed,
depending on the observers' reference frames when
the two events are
not causally conntected. This is analogous to the following fact:
The
order of the degree of entanglement for two
 certain mixed states can be
reversed, depending on the entanglement measures.
A trajectory of quantum states induced by
a  local quantum operation assisted by classical
communications (LOCCs)
 corresponds to a dynamic trajectory of a
particle in special relativity. 

A few authors have proposed serveral entanglement measures, which
quantify the degree of entanglement of quantum states, such as
 negativity
of entanglement $E_n$ \cite{horo,lee,vida}, 
entanglement of cost
$E_C$, entanglement of distillation $E_{D}$ \cite{bene}, 
and quantum
relative entropy of entanglement $E_r$ \cite{vede}. These are
{\it reasonable} measures in the sense that they satisfy the three
necessary conditions \cite{bene,vede,ved2}: (C1) $E(\rho)=0$ if and
only if the density operator $\rho$ of a bipartite system is separable,
(C2) local unitary operations leave $E(\rho)$ invariant, and (C3) the
expected degree of entanglement does not increase under any
LOCCs.
 
The entanglement of quantum states has a rich structure and 
require, in general, multiple parameters for complete 
characterization.
It was shown that a single parameter is sufficient to 
characterize
pure entangled states of bipartite systems in the asymptotic case
\cite{pope,vid2}. 
In this special
case, all entanglement measures are reduced to
a unique measure \cite{rudo,dona}.
 In general, however, the entanglement requires 
multiple parameters for characterization,
for example, $N-1$ parameters for pure entangled states of
bipartite
$N$-dimensional
%($N \times N$) 
systems \cite{vid2,zycz,ben2}.

For better understanding and manipulation of entangled states,
 quantum states must be classified as well as possible.
The degree of entanglement of an entangled pair 
can be compared by investigating the existence of LOCCs that
transform one state to another. One state can be
said to be more entangled
than another if the former can be transformed to the latter by
LOCCs
with unit efficiency. However,
some pairs of entangled states exist such that one cannot be
transformed to the other by any LOCCs. These pairs of states
 cannot be
compared with each other; in other words, they are `incomparable' 
(A simple example is given in Ref. \cite{niel}).

%In non-asymptotic case, on
%the other hand, multiple parameters are necessary 
%It is not known whether bipartite mixed states can be described by a
%single parameter even in the asymptotic case.  

Consideration of orderings in degree of entanglement are important
 in
analyzing the structure of entanglement.
Two certain entanglement measures $E_A$ and $E_B$ are defined
to have the same ordering if they satisfy the following condition for
any two density operators $\rho_i$ and $\rho_j$;
\begin{equation}
\label{a}
E_A(\rho_i) > E_A(\rho_j) \Leftrightarrow
E_B(\rho_i) > E_B(\rho_j).
\end{equation}
Recently, Virmani and Plenio showed that if two certain entanglement
measures, which are
 identical for pure states, give different degrees of entanglement
for some mixed states, the orderings of the two measures are 
different
for those mixed states \cite{virm}.
 For examples, the existence of a bound
entangled state and some losses \cite{vid3,hor2} in purification
processes cause the entanglement of cost $E_C$ to be strictly
greater than entanglement of distillation $E_D$
even in asymptotic cases. (In the finite case,
it was explicitly shown by
numerical analysis that the entanglement of formation
 $E_f$ \cite{bene} and the negativity of entanglement 
$E_n$ give different orderings \cite{eise}.) Thus, the orderings 
 in the degree of entanglement
depend on the measures. This fact suggests that the
various entanglement measures do not have to give the same ordering
for mixed states even in asymptotic cases.

We note that, although the conclusion appears to be odd, it does not
give rise to any bare contradictions. Let us consider two
density operators $\rho_i$ and $\rho_j$. The fact that the
order depends on the entanglement measures 
obviously means that
degree of entanglement of $\rho_i$ is less than that of
 $\rho_j$ in one of the
two measures and vice versa in the other. That is, we have either
\begin{eqnarray}
\label{b}
E_A(\rho_1) > E_A(\rho_2)
\hspace{2mm} \mbox{and} \hspace{2mm}
E_B(\rho_1) < E_B(\rho_2),
\end{eqnarray}
or
\begin{eqnarray}
\label{c}
E_A(\rho_1) < E_A(\rho_2)
\hspace{2mm} \mbox{and} \hspace{2mm}
E_B(\rho_1) > E_B(\rho_2).
\end{eqnarray}
A quantum state with density operator $\rho_i$ cannot be
transformed to one with $\rho_j$ by any LOCCs due to 
condition (C3)
 if the degree of entanglement of $\rho_i$ is less than $\rho_j$ in
any of the reasonable entanglement measures. 
(With less efficiency than
one, forbidden paths may be allowed by 
LOCCs.) Thus, the pairs of states are
incomparable \cite{niel}, implying
the necessity of multiple parameters for 
characterizing the entanglement of bipartite mixed states.
Now, it is interesting to note that this fact is analogous to a fact
of the
special relativity.  The temporal order of two certain events 
depends on
observers' reference frames \cite{tayl}.  Although it appears to be
odd, this fact does not give rise to contradictions because the two
events cannot be causally connected (or 
can only be space-likely connected) in this case. 
 Let us consider a map where each
mixed state is coordinated by degree of entanglement of two certain
measures, $E_A$ and $E_B$. (See Fig. 1.)
 We can easily see that a point in the map
cannot be moved by any local unitary operation due to
condition (C2). Thus, a class of mixed states,
 which are equivalent within a local
unitary operation, corresponds to a point in the map.  In general, the
converse is not true.  The larger the number of reasonable
entanglement measures we adopt, the more refined the 
coordination of
density operators will become.  If LOCCs are applied,
 a point can flow
 through a trajectory that always points in the lower-left direction
from a point in the map. This fact is analogous to a fact in
special relativity; each observer goes through a path in
space-time that always points from a point
 to somewhere within the light cone.
  Here, a point and a trajectory in the map, respectively,
correspond to an event and an observer's path in space-time.  
That there is no unique entanglement measure 
is also analogous to there being no preferred reference frame in 
special relativity. 
Multiplicity of
measures does not mean oddity, but a higher dimensional structure of
degree of entanglement of quantum states. 
 In fact, the conclusion that
multiple parameters are needed has been logically obtained by 
Vidal
and by Bennett {\it et al.}  in the case of finite pure
states where incomparable states exist \cite{vid2,ben2}. 
What we have shown here is that existence of incomparable states and
multiple measures are in close analogy with some facts in the special
theory of relativity.

%The analogy to special relativity suggests several open questions.

In conclusion, various entanglement measures for quantum bipartite
mixed states do not give the same ordering in general \cite{virm},
even in asmyptotic cases.
That is, two certain mixed states' ordering in the degree of
 entanglement
depends on the entanglement measures.  However, 
such pairs of mixed
states cannot be transformed to each other by LOCCs,
so they are incomparble.  We make an analogy between these facts
 and the
relativity of temporal order in the special theory of relativity.
 The non-uniqueness of the entanglement measure is also in
 analogy with the
non-existence of a preferred reference frame. 
Our hope is that this analogy
will inspire other ideas which may lead us to a better 
understanding of the structure
of entanglement measures. It is notable that after
this work the analogy was 
persued in the case of
finite pure-state entanglement \cite{zycz}. 

\acknowledgments 
This work was supported by the Korean Ministry of
Science and Technology through the Creative Research Initiatives
Program under Contract No. 99-C-CT-01-C-35.

Fig. 1;
Consider a map where each
mixed state is coordinated by the degrees of entanglement of two
certain measures, $E_A$ and $E_B$. 
If LOCCs are applied, a point can flow 
through a trajectory that always points in the lower-left direction
from a point in the map. This fact is in analogy with 
special relativity where each observer goes through a path in
space-time that always points to somewhere within the
 light-cone from a
point.  A point and a trajectory in the map, respectively,
correspond to an event and an observer's path in space-time.  
{\it The states corresponding to $p$ and $q$ are incomparable 
since they
cannot be transformed, with unit efficiency, to each other
by any LOCCs.}
\end{document}